\documentstyle[12pt,titlepage]{article}

\renewcommand{\theequation}{\thesection.\arabic{equation}}

\font\grande=cmr10 scaled \magstep4
\font\medio=cmr10 scaled \magstep2
\outer\def\beginsection#1\par{\medbreak\bigskip
      \message{#1}\leftline{\bf#1}\nobreak\medskip
\vskip-\parskip
      \noindent}

\def\laq{\raise 0.4ex\hbox{$<$}\kern -0.8em\lower 0.62
ex\hbox{$\sim$}}
\def\gaq{\raise 0.4ex\hbox{$>$}\kern -0.7em\lower 0.62
ex\hbox{$\sim$}}

\def\beq{\begin{equation}}
\def\eeq{\end{equation}}
\def\bea{\begin{eqnarray}}
\def\eea{\end{eqnarray}}

\def \ap {\alpha^{\prime}}

\begin{document}
\bibliographystyle {unsrt}

\titlepage
\begin{flushright}
March 21 1997 \\
CERN-TH/97-42 \\
hep-th/9703150

\end{flushright}
\vspace{10mm}
\begin{center}
{\grande Inhomogeneous Pre-Big Bang String Cosmology}\\

\vspace{10mm}

G. Veneziano \\
{\em Theory Division, CERN, CH-1211 Geneva 23, Switzerland} \\
\end{center}
\vspace{10mm}
\centerline{\medio  Abstract}

\noindent
An inhomogeneous version  of
pre--Big Bang cosmology emerges, within string theory, from quite generic
 initial conditions, provided they lie deeply inside the
weak-coupling, low-curvature regime. Large-scale homogeneity, flatness, and
isotropy appear  naturally  as late-time outcomes of such an evolution.
 \vfill
\begin{flushleft}
CERN-TH/97-42 \\
March 1997
\end{flushleft}

\newpage

\renewcommand{\theequation}{1.\arabic{equation}}
\setcounter{equation}{0}
\section {Introduction}
 Superstring theory, a supposedly consistent and unified quantum theory of
all interactions, suggests, through its peculiar duality symmetries
\cite{duality},
 an attractive alternative to standard inflationary cosmology, the
so-called pre- Big Bang (PBB) scenario
  \cite{GV1} \cite{MG1}
\cite {MG2}.

The basic postulate of  PBB
cosmology is that the Universe started its evolution from a very
perturbative initial state, i.e. from very weak coupling and very small
curvatures (in appropriate units, see below). If initial homogeneity is
also assumed, this leads automatically
 to a super-inflationary epoch during
which the accelerated expansion of the Universe is driven by the
accelerated growth of the effective (gauge and gravity) coupling, i.e. of
the dilaton.

This result looks surprising, at first sight, since the equation of
state of the dilaton is not of the kind that one usually associates  
with inflation. Indeed, inflation only occurs if the geometry is  
looked at
in the so-called String(S)-frame, the one appearing in the
Nambu-Goto-Polyakov action, while in the  Einstein(E)-frame one  rather
observes an accelerated contraction. The point here is that,
precisely at weak coupling and small curvatures, fundamental strings keep
their characteristic size constant and sweep geodesic surfaces with respect
to the S-frame metric \cite{SV} \cite{CLO}. Thus, in this regime at least,
the latter {\it is} the correct frame for describing how distances
among objects vary w.r.t. their intrinsic size, and for
determining what is actually meant by small or large curvatures. As long as
curvatures remain small (resp. become $O({1\over \ap})$) in the string
frame, the low energy effective action of string theory represents an
adequate (resp. becomes an inadequate) description of physics.

Going from one ``frame" to another, although not
 a coordinate transformation,
 is only a matter of field redefinitions. Hence, any physical
observable takes the same value in any such ``frame". Whenever useful for
computations, any other frame can be used, and we shall take  advantage of
this possibility below. However, the description of physical phenomena is
all but intuitive in other frames and, in order to avoid problems of
interpretation, it is best to transform back the results to the string
frame at the end of the calculations.

The PBB scenario also needs an ``exit assumption" from the PBB phase to
standard non-inflationary cosmology. Higher derivative ($\ap$) corrections
are expected to stop the indefinite growth of curvature
(see \cite{ex} for recent progress on
this issue), while
 loop corrections and/or a non-perturbative potential are usually assumed
to stop the indefinite growth of the coupling itself. If such a ``graceful
exit" turns out to be a property of string theory, one will obtain an
interesting scenario whereby the
usually assumed Big Bang initial conditions (a hot, dense and
highly-curved state) will be the outcome -rather than the starting point- of
inflation. Several interesting phenomenological consequences of
the scheme have also been worked out \cite{phen}.

In spite of the above advantages of the PBB scenario, it is clear that the
assumption of homogeneous initial conditions is a difficult one to swallow.
After all, inflation is supposed to explain \cite{inflation}  
homogeneity(as well as isotropy and flatness)! The purpose of this  
note is to relax
the assumption of initial homogeneity by taking  the initial state of the
Universe to be in the weak-coupling, weak-curvature regime, but otherwise
arbitrary.
Generically, PBB behaviour will emerge in suitable regions of
space which, eventually, will fill all but an infinitesimal fraction  
ofspace. Within those regions, the Universe will appear very  
homogeneous,
isotropic and spatially flat.

The plan of the paper is as follows: in Section 2, I give the general
form of the equations in the E-frame and describe some general
features of their solutions, which will be
confirmed by the construction, in Section 3, of the general
 analytic asymptotic solution near the  ``singularity". In
Section 4,  the results of the previous Sections will be reinterpreted in
the physical S-frame and semiquantitative local
conditions for the onset of PBB behaviour will be given.

Most of our technical results are not really new, only
 the physical motivation
and interpretation are. In particular, our method can be traced back to the
work of Belinskii and Kalatnikov \cite{LK} which was made recently more
systematic through the so-called gradient expansion  method either
directly in terms of Einstein's equations \cite{ge}, or within a
Hamilton-Jacobi formulation \cite{HJ}.

 \renewcommand{\theequation}{2.\arabic{equation}} \setcounter{equation}{0}
\section {General properties of the solutions in the E-frame}
Having made the basic postulate that the Universe started its (pre)
history in a weak-coupling, low-curvature state, we will describe its
early evolution in terms of the tree-level,
low-energy effective action of
string theory.

For simplicity we shall consider only the particular case of a
critical superstring theory, with vanishing cosmological constant and
six frozen internal dimensions. The effective four-dimensional theory is
thus described (in Landau-Lifshitz
notations \cite{LL}) by the action:
\beq
\Gamma^{S}_{eff} = {1 \over 2} \int dx \sqrt{G} e^{-\phi}\left(R(G) +
G^{\mu\nu}
 \partial_{\mu} \phi \partial_{\nu} \phi\right)
\label{Saction}
\eeq
where $G_{\mu\nu}$ is the string-frame (S-frame) metric, $\phi$ is the
dilaton and we have also set to zero the
antisymmetric tensor field $B_{\mu\nu}$.  In the conformally related
Einstein frame (E-frame) defined by the new metric $g_{\mu\nu}$ via:
 \beq
 g_{\mu\nu} = e^{-\phi} G_{\mu\nu}
\label{ESframe}
\eeq
 the action becomes:
\beq
\Gamma^{E}_{eff} = {1 \over 2} \int dx \sqrt{g} \left(R(g) -
{1 \over 2} g^{\mu\nu}
 \partial_{\mu} \phi \partial_{\nu} \phi\right)
\label{Eaction}
\eeq
and the field equations take  the familiar form:
\bea
R_{\mu \nu} - {1 \over 2} g_{\mu \nu} R = {1 \over 2}
 \partial_{\mu} \phi \partial_{\nu} \phi -{1 \over 4}  g_{\mu\nu}
 (\partial \phi)^2 \nonumber \\
\bigtriangledown^2 \phi \equiv g^{\mu \nu}  D_{\mu} \partial_{\nu} \phi = 0
\eea
where we are using units in which $8 \pi G_N =1$.

We also choose to work in the synchronous gauge, i.e. we fix:
\beq
g_{00} = -1, \; g_{0i} = 0 \; (i= 1,2,3),
\eeq
so that the dynamical variables become the three-metric $g_{ij}$ and the
dilaton $\phi$.

The Einstein equations for the system can be readily written. They  
consist of:
\begin{itemize}
\item  The Hamiltonian constraint
\beq
2 \dot{\phi}^2 + Tr\chi^2 - \chi^2 = 4 {\cal R} - 2 (\bigtriangledown
\phi)^2 \label{HC}
\eeq
\item The three Momentum Constraints
\beq
\chi_{,i} - \chi_{i;j}^j = - \dot{\phi}\bigtriangledown_i \phi
\label{PC}
\eeq
\item  The dynamical equation for $\phi$
\beq
\ddot{\phi} +{1 \over 2} \chi \dot{\phi} = \bigtriangledown^2 \phi
\label{PhiE}
\eeq
\item The  dynamical equation for $g_{ij}$
\beq
\dot{\chi}_i^j +{1 \over 2} \; \chi \;\chi_i^j = -2 {\cal R}_i^j +
 \bigtriangledown_i
\phi \bigtriangledown^j \phi
\label{gE}
\eeq
\end{itemize}
where a dot denotes derivative w.r.t. E-frame ``cosmic" time $\tau$,
covariant derivatives are defined w.r.t. the three-metric $g_{ij}$ (whose
three-curvature is denoted by $\cal{R}$), and we have defined:
\beq
\chi_{ij} = \dot{g_{ij}}, \;  \chi_i^j = \chi_{ik}g^{kj}, \;
Tr\chi^2 = \chi_i^j \chi_j^i, \; \chi = \chi_i^i = {\dot{g} \over g}, \; g =
det(g_{ij}).
 \label{chidef}
\eeq
It is easy to check that the above equations reduce to the usual
cosmological equations once all spatial gradients are set to zero. It is
also possible to check that the four constraints, once imposed at an
initial time, remain valid at all times thanks to the evolution equations.

The following general inequalities follow as long as the three metric  
ispositive (semi) definite:
\beq
 Tr\chi^2 \ge {1 \over 3}  \chi^2 \ge 0 \; ,
\label{posbounds}
\eeq
where the first equality is reached in the isotropic case.

Combining the trace of (\ref{gE}) and (\ref{HC}) we obtain
a useful equation
which does not involve spatial gradients:
\beq
\dot{\chi} = - \left ({1 \over 2} Tr\chi^2 +  \dot{\phi}^2 \right) \le 0 ,
\label{CombE}
\eeq
with the equality sign holding iff $\dot{\phi} = \chi_{ij} =0$.

Equation (\ref{CombE}), together with the basic postulate of PBB cosmology,
is enough to conclude that, on the
$\tau =0$ hypersurface, we can only have regions  with
 $\chi<0$. Indeed, regions with $\chi>0$, given Eq. (\ref{CombE}), will
 necessarily have to originate from a large curvature phase in the past.
Even regions with  $\chi=0$ are excluded, since, going backward in time,
they generically came from a positive $\chi$. Furthermore, the Hamiltonian
constraint (\ref{HC}) implies  that $\chi$ can only vanish if
 \beq
 {\cal R} \ge   {1 \over 4} Tr\chi^2  + {1 \over 2} \dot{\phi}^2 ,
 \label{RBound}
\eeq
i.e. in regions of sufficiently large positive spatial curvature. In
sufficiently homogeneous regions, in which fields vary little over a Hubble
radius, we get from (\ref{posbounds}), (\ref{HC}) the  bounds:
\beq
 - \sqrt{3} \le \rho \equiv {\chi \over \sqrt{ Tr\chi^2}} \le -1 .
\label{rhoBound}
\eeq

We conclude that generic initial data satisfying the PBB postulate can be
presented in terms of quasi-homogeneous regions where
$-\sqrt{3}<\rho<-1$,  separated by inhomogeneous
regions where $\rho$ {\it can} take any negative value larger than
$-\sqrt{3}$.  Let us
see how each one of these regions  evolves in time:

The quasi-homogeneous regions with $\rho<0$ undergo an accelerated
contraction in the Einstein frame  and are consistent with the PBB
postulate. We shall see below (and in Section 3) more precisely  how they
evolve in the E-frame while, in Section 4, we will reinterpret those results
in the S-frame.

Highly inhomogeneous regions are more difficult to deal with, but are not
expected \cite{LK} \cite{ge} \cite{HJ} to undergo a marked accelerated (or
decelerated) evolution.

It is  relatively easy to see that, for quasi-homogeneous regions with
$\rho<0$,
the approximation of neglecting spatial derivatives improves with  
time.Leaving to  Section 3 the study of the
 asymptotic solution,
we just note that, neglecting spatial gradients, the trace of (\ref{gE})
gives: \beq
\chi \sim {2 \over \tau - \tau_0} \Rightarrow g \sim (\tau - \tau_0)^2 ,
\tau < \tau_0 ,
\label{solchi}
\eeq
where, in general, $\tau_0$ can depend on $x$. Invoking now (\ref{HC}) and
the definition (\ref{rhoBound}) we can determine $\dot{\phi}$ up to a sign
ambiguity. However, given the fact that we chose $\chi<0$, and that we want
to start the evolution from weak coupling, we have to choose the sign
giving:
  \beq \dot{\phi} = - {\chi \over \sqrt2}
\sqrt{1-\rho^{-2}}  \Rightarrow \phi \sim \; - c(x) log (\tau_0 -
\tau), \; \; c(x) >0.
\label{phidot}
 \eeq

In order to check  that spatial gradients are subleading
w.r.t. time derivatives as we approach $\tau = \tau_0$ consider, for
instance, the ratio (with no sum over $i,j$ implied):
\beq
r_{ij} \equiv {g^{ij} \bigtriangledown_i
\phi \bigtriangledown_j \phi \over \dot{\phi}^{2}}.
\label{grads}
\eeq
As we approach the singularity, using (\ref{solchi}), (\ref{phidot}),
we find:
\beq
r_{ij} \sim g \cdot g^{ij} \bigtriangledown_i
\phi \bigtriangledown_j \phi  \sim   \epsilon^{ikn} \epsilon^{jlm} g_{kl}
g_{nm} \; log^2 (\tau_0 - \tau) \rightarrow 0,
\eeq
unless the contraction of the three-volume is highly anisotropic.

We thus conclude that, within sufficiently homogeneous and
isotropic regions, the approximation of neglecting spatial derivatives
becomes increasingly good as we approach the singularity at $\tau =
\tau_0$. This conclusion will be fully confirmed in the analytic asymptotic
solution presented in  Section 3.

Let us now try to understand how things evolve as one goes backward in
time, $\tau \rightarrow -\infty$. It is not possible to
obtain an asymptotic solution in this limit by the method used in Section 3.
Nevertheless, we can make the following observations: Eqs. (\ref{PhiE}),
(\ref{CombE}), together imply
 that the only early-time fixed points are either at
infinity (corresponding to a singularity)  or at
$\dot{\phi} = \chi_{ij}= \bigtriangledown^2 \phi=0$.
 This implies however (neglecting boundary terms) $\bigtriangledown_i
\phi=0$ and thus, using Eq. (\ref{gE}),  ${\cal R}_i^j=0$, which,
 in three dimensions, is equivalent to flat space. Thus the only
non-singular fixed point is the trivial one.

 Homogeneous PBB cosmologies are
examples of solutions converging towards
 this trivial fixed point in the far
past. Our basic postulate requires that there should be a {\it
finite-measure} basin of attraction (containing the homogeneous cases)
towards this fixed point. By using a reliable linear perturbation theory
around the homogeneous solutions, it is not hard to show that such a finite
basin should exist, although determining its exact extension is highly
non-trivial.

 Another related issue is whether, within such a basin of attraction, one
is really approaching the trivial Minkowski vacuum with
vanishing coupling and curvature. The tricky point here is that
curvatures and derivatives of $\phi$ have to go to zero in {\it string}
units, namely they have to become much smaller than $e^{\phi}$  in the
E-frame. Although we have indications that this is indeed the case,
we leave a full discussion of this important question to further
investigations.

\renewcommand{\theequation}{3.\arabic{equation}}
\setcounter{equation}{0}
\section {General asymptotic solution in the E-frame}

In this Section we present the leading term of the general asymptotic
solution of Eqs. (\ref{HC})-(\ref{gE}) near the singularity hypersurface
$\tau = \tau_0(x)$. The construction is quite straightforward (see e.g.
(\cite{LK})), if one makes a few approximations in solving the equations
and then checks, a posteriori, their validity. Finally, a counting of the
number of arbitrary functions contained in the solution shows
that we are indeed describing the most general solution.

 Inserting (\ref{solchi}) back
into Eq.(\ref{gE}) we get, for small enough gradients,
\beq
\dot{\chi}_i^j + {1 \over \tau - \tau_0} \chi_i^j = 0, \tau < \tau_0 .
\eeq
Integrating  this equation twice we get:
\beq
\chi_i^j = { 2 \lambda_i^j \over \tau - \tau_0}, \;
 g_{ij}(x,\tau) =
\left[ exp \left(2 \lambda(x) log (1- {\tau  \over  \tau_0})\right) \right
] _i^k \;
 g_{kj}(x,0)
\label{gsol}
\eeq
where the space-dependent quantities $\lambda_i^k$ satisfy:
\beq
\lambda_i^i =1, \; {1 \over 3} \le \lambda_i^k \lambda_k^i \le 1, \;
  \lambda_{ij} \equiv  \lambda_i^k g_{kj}(x,0) = \lambda_{ji}.
\eeq
At this point we use (\ref{HC}) and find, asymptotically:
\beq
\phi(x,\tau) = \phi(x,0) - \sqrt{2} \sqrt{1- \lambda_i^k \lambda_k^i}\;
 log (1- {\tau  \over  \tau_0}).
\label{phisol}
\eeq
Rewriting finally Eq.(\ref{PhiE}) in the form:
\beq
(\sqrt{g} \dot{\phi})^. = \sqrt{g}\bigtriangledown^2 \phi,
\eeq
we see that it is automatically fulfilled asymptotically since, from
(\ref{gsol}), (\ref{phisol}),  we find $\sqrt{g} \dot{\phi} \sim
const(x)$.

A convenient way to present the solution and to check that, indeed, it is
general, consists of introducing local coordinates with respect to which
the matrix $\lambda$ and the three-metric are diagonal. Then we can write:
\bea
g_{ij}(x,\tau) = \sum_a e_i^a(x) e_j^a(x)
(1- {\tau  \over  \tau_0})^{2\lambda_a(x)} \\
\phi(x,\tau) = \phi(x,0) - \sqrt{2} \sqrt{1- \sum \lambda_a^2}\;
 log (1- {\tau  \over  \tau_0}), \;  \sum \lambda_a = 1
\eea
where, at the moment, $e_i^a$ are arbitrary ``dreibein" matrices and we note
that the parameter $\rho$ of  Section 2 is now given by
$\rho^{-2} = \sum \lambda_a^2$.

The above solution appears to depend upon thirteen arbitrary functions
of space, i.e. nine $e_i^a$, two of the three $\lambda_a$, $\tau_0$,
and $\phi_0$. However, we have not imposed yet the momentum constraints
(\ref{PC}). These constraints, being conserved,  reduce to
their form at $\tau=0$ and  can be used to eliminate three of the nine
$e_i^a$. We are left with ten arbitrary functions. Yet, within the
synchronous gauge, we can still perform a four-parameter family of gauge
transformations (see, e.g.,  \cite{LL}) allowing us to set $\tau_0(x) =
\tau_0$ (independent of $x$) and to further fix three of the  
remaining sixindependent quantities $e_i^a$. We end up with six  
(physical) arbitrary
functions which is, indeed, the correct number \cite{LK} for the problem at
hand.

\renewcommand{\theequation}{4.\arabic{equation}}
\setcounter{equation}{0}
\section { Physical reinterpretation in the string frame }

In order to discuss the physical implications of our results we transform
them back to the string frame.  Using Eq. (\ref{ESframe})
 we obtain
\beq
G_{ij}(x,\tau) = \sum_a e_i^a(x) e_j^a(x) \cdot e^{\phi(x,0)}
(1- {\tau  \over  \tau_0})^{2\lambda_a - \sqrt{2} \sqrt{1- \sum
\lambda_b^2}}  \; \; ,
\eeq

Since $g_{00}= -1$ has also been changed into a non-trivial $G_{00}$,  
westill have to go to the synchronous gauge in the S-frame. This is
easily accomplished, to leading order in the gradients,
 by changing the time
variable from $\tau$ to $t$ according to:
\beq
\sqrt{2} (t_0 - t)   = \tau_0 \; e^{\phi(x,0)/2}
(\sqrt{2} - \sqrt{1 -
\sum \lambda_a^2 })^{-1}
(1- \tau / \tau_0)^{1-{1\over \sqrt{2}} \sqrt{1- \sum
\lambda_a^2}}
\eeq
where we note that, for any choice of the $\lambda_a$, $t \rightarrow t_0$
as $\tau  \rightarrow  \tau_0$. Finally, after defining approriate
S-frame
 ``dreibeins" $E_i^a(x)$, we get:
\bea
G_{ij}(x,t) = \sum_a E_i^a(x) E_j^a(x)
(1- {t  \over  t_0})^{2\alpha_a(x)} \\
\phi(x,t) = \phi(x,0) + \gamma \;
 log (1- {t  \over  t_0}),
\eea
where:
\beq
\alpha_a = {\lambda_a - {1\over \sqrt{2}} \sqrt{1- \sum
\lambda_b^2} \over 1- {1\over \sqrt{2}} \sqrt{1- \sum
\lambda_b^2}},\;
\gamma = - {\sqrt{2} \sqrt{1- \sum
\lambda_b^2} \over 1- {1\over \sqrt{2}} \sqrt{1- \sum
\lambda_b^2}}.
\label{alphas}
\eeq
 The above relations imply:
\beq
\sum \alpha_a^2 = 1, \; \gamma = -1 + \sum \alpha_a\;
\eeq
so that we find for the ($O(d,d)$-invariant) shifted dilaton
\cite{duality}:
 \beq
 \bar{\phi} \equiv \phi
- {1 \over 2} tr log G  \sim  - log (1- {t  \over  t_0}) + \bar{\phi}(x,0),
\eeq
which generalizes the well known S-frame homogeneous cosmology results
\cite{GV1}, \cite{MG1}.

Eqs. (\ref{alphas}) can  easily be inverted to express the $\lambda_a$'s in
terms of the $\alpha_a$'s:
\beq
\lambda_a = {1 \over 3} +{2 \over 3} {\alpha_a - {1 \over 3} \sum
\alpha_b \over 1 - {1 \over 3} \sum
\alpha_b }.
\eeq

Using these equations we can construct, in principle,
scale-factor-duality(SFD) related solutions ($\alpha_a \rightarrow -
\alpha_a$ for some $a$\cite{GV1}) even in the inhomogeneous case. Of course,
one has to check that the constraints can  be simultaneously imposed on each
component of the dual pairs.

 We are
finally in a position to study how the three-volume of a quasi-homogeneous
region evolves in the string frame. We find
 \beq
{1 \over 2} \chi^{(s)} \equiv {1 \over 2} {\dot{G} \over G} =
- {\sum \alpha_a \over t_0 - t}
\label{chis}
\eeq
which shows that  quasi-homogeneous regions undergo
{\it superinflation} in the string frame provided:
\beq
\sum \alpha_a = {1 - {3 \over \sqrt{2}} \sqrt{1- \sum \lambda_a^2} \over
1 - {1\over \sqrt{2}} \sqrt{1- \sum
\lambda_a^2}} <0.
\label{cond}
\eeq
This condition is satisfied provided $ \rho^{-2} = \sum \lambda_a^2 < 7/9$,
which includes all but a small region for the allowed values of this
quantity (see (\ref{rhoBound})). Note that the maximal expansion rate is
reached for $\sum \alpha_a = - \sqrt{3}$  corresponding to the isotropic
case $\lambda_a = 1/3$.

We may also ask about the flatness problem. It is easy to see that the
typical quantity measuring flatness, ${{\cal R} \over \chi^{2}}$, contains
a logarithmically enhanced term which, for $t \rightarrow t_0$,  
behaves as\beq
{{\cal R} \over \chi^{2}} \sim {G^{ij} \bigtriangledown_i
\phi \bigtriangledown_j \phi \over \dot{\phi}^{2}}
\sim (t-t_0)^{2-2 {\rm Max}(\alpha_a)} log^2 (t_0 - t) ,
\eeq
 and thus goes to zero almost everywhere thanks to
(4.6). The non-enhanced terms are more involved, but certainly behave in
the same way for sufficiently isotropic situations. Thus, as expected,
spatial flatness is generically achieved after a long PBB era.

We conclude that all but the ``skin" of the quasi-homogeneous regions
with negative $\rho$ inflates when we look at the geometry in the ``right"
frame and becomes homogeneous and spatially flat. Presumably,
after a while, these regions represent by far the largest fraction of the
Universe. Furthermore, the isotropic regions dominate  all others,
providing also a possible ``explanation" for the isotropy and flatness of
our observable Universe.

\renewcommand{\theequation}{6.\arabic{equation}}
\setcounter{equation}{0}
\section {Conclusions}

In this paper we have relaxed the assumption of homogeneity (and
spatial flatness)
in the initial conditions for string cosmology. The emerging picture,
which bears some similarity to Linde's chaotic inflation \cite{Linde}, can
be described as follows: the
primordial state of the Universe is  arbitrarily close to flat space-time
but not particularly homogeneous, in the sense that  time and
space derivatives are both extremely small but comparable.
Stochastically, however, sufficiently
homogeneous and isotropic conditions will emerge in some regions of space
and undergo, in the S-frame, an accelerated expansion yielding
homogeneity, flatness and isotropy as one approaches a singularity in the
future within a finite proper time. Before such a
singularity is reached, the low-energy effective action will cease to be
valid and, hopefully, the higher-derivative corrections
 present in string theory will lead  to a  finite-curvature
stringy phase \cite{ex}
rather than to a genuine singularity.

In this connection, it is amusing to speculate that the
famous singularity theorems of Hawking and Penrose \cite{HP} will find an
unexpected (and welcome!) application in string cosmology: since the
conditions of the theorems do probably apply to the low-energy effective
action, they will imply the necessary growth of curvature up to  the scale
at which string corrections invalidate the assumptions made in those
theorems, i.e. the unavoidable occurrence of a long, dilaton-driven
inflationary phase as long as the initial Universe was indeed at very weak
coupling and very nearly flat.

\vskip .5 cm

\section*{Acknowledgements}

I am grateful to R. Brustein, A. Buonanno, K.A. Meissner,
 M. Gasperini and
C. Ungarelli for many useful comments on the material presented here. This
work was supported in part by the EC contract No. ERBCHRX-CT94-0488.

 \end{document}